\documentclass[12pt,a4paper,dvips]{article}
\usepackage{a4p}
\usepackage{cite,mcite}
\usepackage{graphicx}
\usepackage{l3_title}
\usepackage{ifthen}
\usepackage{Lep}
\usepackage{amsmath}
\usepackage{amssymb}
\usepackage{epsfig}

\journalname{Phys. Lett. B}
\date{August 05, 1999}
\preprint{99-116}
\Lep{1}

\newlength{\capindent}
\newlength{\capwidth}
\newlength{\figwidth}
\newcommand{\icaption}[2][!*!,!]{\hspace*{\capindent}%
  \begin{minipage}{\capwidth}
    \ifthenelse{\equal{#1}{!*!,!}}%
      {\caption{#2}}%
      {\caption[#1]{#2}}
  \end{minipage}}

\newcommand{\etal}{\textit{et~al.}}
\newcommand{\pho}{\phantom{0}}
\newcommand{\stat}{\ensuremath{~(\text{stat})}}
\newcommand{\syst}{\ensuremath{~(\text{syst})}}

\newcommand{\mrad}{\ensuremath{\mathrm{mrad}}}
\newcommand{\MeV}{\ensuremath{\mathrm{Me\hspace{-0.1em}V}}}
\newcommand{\GeV}{\ensuremath{\mathrm{Ge\hspace{-0.1em}V}}}
\newcommand{\pb}{\ensuremath{\mathrm{pb^{-1}}}}
\newcommand{\Br}{\ensuremath{\mathrm{Br}}}
\newcommand{\jq}{\ensuremath{j_{\quark}}}
\newcommand{\Zboson}{\ensuremath{\mathrm{Z}}}
\newcommand{\electron}{\ensuremath{\mathrm{e}}}
\newcommand{\quark}{\ensuremath{\mathrm{q}}}
\newcommand{\Quark}{\ensuremath{\mathrm{Q}}}
\newcommand{\dquark}{\ensuremath{\mathrm{d}}}
\newcommand{\uquark}{\ensuremath{\mathrm{u}}}
\newcommand{\squark}{\ensuremath{\mathrm{s}}}
\newcommand{\cquark}{\ensuremath{\mathrm{c}}}
\newcommand{\bquark}{\ensuremath{\mathrm{b}}}

\newcommand{\Bmeson}{\ensuremath{\mathrm{B}}}
\newcommand{\BmesonUD}{\ensuremath{\mathrm{B}_{\uquark,\dquark}}}
\newcommand{\Bstar}{\ensuremath{\mathrm{B}^{*}}}

\newcommand{\BorBstar}{\ensuremath{\mathrm{B}^{(*)}}}
\newcommand{\Bdstar}{\ensuremath{\mathrm{B}^{**}}}
\newcommand{\BdstarUD}{\ensuremath{\mathrm{B}^{**}_{\uquark,\dquark}}}
\newcommand{\BdstarS}{\ensuremath{\mathrm{B}^{**}_{\squark}}}
\newcommand{\Btwostar}{\ensuremath{\mathrm{B}_2^{*}}}
\newcommand{\Bone}{\ensuremath{\mathrm{B}_1}}
\newcommand{\Bonestar}{\ensuremath{\mathrm{B}_1^{*}}}
\newcommand{\Bzerostar}{\ensuremath{\mathrm{B}_0^{*}}}
\newcommand{\Bprime}{\ensuremath{\Bmeson'}}
\newcommand{\Dmeson}{\ensuremath{\mathrm{D}}}

\newcommand{\Dstar}{\ensuremath{\mathrm{D}^{*}}}

\newcommand{\Kmeson}{\ensuremath{\mathrm{K}}}

\newcommand{\Rb}{\ensuremath{R_{\bquark}}}

\begin{document}


\begin{titlepage}
\title{Measurement of the Spectroscopy of Orbitally Excited B~Mesons at LEP}
\author{The L3 Collaboration}


\begin{abstract}
We measure the masses, decay widths and relative production rate of orbitally
excited B mesons using $1.25$ million hadronic Z decays recorded by the L3
detector.  B-meson candidates are inclusively reconstructed and combined with
charged pions produced at the primary event vertex.  An excess of events above
the expected background in the $\Bmeson\pi$ mass spectrum in the region
$5.6-5.8~\GeV$ is interpreted as resulting from the decay
$\BdstarUD\rightarrow\BorBstar\pi$, where $\BdstarUD$ denotes a mixture of $l=1$
B-meson states containing a $\uquark$ or a $\dquark$ quark.  A fit to the mass
spectrum yields the masses and decay widths of the $\Bonestar$ and $\Btwostar$
spin states, as well as the branching fraction for the combination of $l=1$
states.  In addition, evidence is presented for the existence of an excited
B-meson state or mixture of states in the region $5.9-6.0~\GeV$.
\end{abstract}

\vspace*{\fill}
\submitted

\end{titlepage}


\section{Introduction \label{sec:Introduction}}

The spectroscopy of orbitally excited B mesons provides important information
regarding the underlying QCD potential.  A flavor-spin symmetry \cite{HQS}
arises from the fact that the mass of the $\bquark$ quark is large relative to
$\Lambda_{\mathrm{QCD}}$.  In this approximation, the spin ($\vec{s}_{\Quark}$)
of the heavy quark ($\Quark$) is conserved in production and decay processes
independently of the total angular momentum
($\vec{j}_{\quark}=\vec{s}_{\quark}+\vec{l}$) of the light quark ($\quark$).
Excitation energy levels are thus degenerate doublets in total spin and can be
expressed in terms of the spin-parity of the meson, $J^P$, and the total spin of
the light quark, $\jq$.  The $l=0$ mesons, for which $\jq=1/2$, have two
possible spin states: a pseudo-scalar $P$, corresponding to $J^P=0^-$, and a
vector $V$, corresponding to $J^P=1^-$.  If the spin of the heavy quark is
independently conserved, the relative production rate of these states would be
$V/(V+P)=0.75$.\footnote{Corrections due to the decay of higher excited states
are predicted to be small.} Recent measurements of the $\Bstar$ production rate
using $\Bstar\rightarrow\Bmeson\gamma$ decays in $\electron^{+}\electron^{-}$
collisions \cite{BstarL3,BstarLEP} agree well with this number.

For orbitally excited mesons with $l=1$, there are two sets of degenerate
doublets: one corresponding to $\jq=1/2$ and the other to $\jq=3/2$.  For the
$\jq=1/2$ doublet, there is one state corresponding to $J^P=0^+$ and three
degenerate states corresponding to $J^P=1^+$.  For the $\jq=3/2$ doublet, there
are three degenerate states corresponding to $J^P=1^+$ and five others
corresponding to $J^P=2^+$.  Rules for the decay of the $l=1$ states to the
$l=0$ states are dictated by spin-parity conservation \cite{SpinParity}.  For
the dominant two-body decays to a pion and a B meson, the $\jq=1/2$ states
undergo an $L=0$ transition (S-wave) and their decay widths are expected to be
broad in comparison to those of the $\jq=3/2$ states which undergo an $L=2$
transition (D-wave).  Table~\ref{tab:decays} presents the nomenclature of the
$l=1$ B mesons containing either a $\uquark$ or a $\dquark$ quark, $\BdstarUD$,
with the corresponding spin states, degeneracies and two-body decay modes.  The
spectroscopy of $l=1$ B mesons containing an $\squark$ quark is not studied in
this analysis, but is examined as a possible source of systematic uncertainty.

\begin{table}[!htb]
  \begin{center}
    \begin{tabular}{|c|c|c|l|c|}
      \hline
      $\jq$ & $J^P$ & $2J+1$ & Decay Mode & Transition \\
      \hline \hline
      $1/2$ & $0^+$ & 1 & $\Bzerostar\rightarrow\Bmeson\pi$          & S-wave \\
      $1/2$ & $1^+$ & 3 & $\Bonestar\rightarrow\Bstar\pi$            & S-wave \\
      $3/2$ & $1^+$ & 3 & $\Bone\rightarrow\Bstar\pi$                & D-wave \\
      $3/2$ & $2^+$ & 5 & $\Btwostar\rightarrow\Bstar\pi,\Bmeson\pi$ & D-wave \\
      \hline
    \end{tabular}
    \icaption{Spin states of the $l=1$ $\BmesonUD$ mesons with the expected
      relative production rates according to spin counting ($2J+1$) and the
      associated decay modes and transitions predicted by spin-parity
      conservation.
      \label{tab:decays}}
  \end{center}
\end{table}

Predictions for the masses and decay widths of the four spin states are based on
Heavy Quark Effective Theory (HQET), in which corrections to the spectator quark
model are expressed as perturbations in powers of
$\Lambda_{\mathrm{QCD}}/m_{\Quark}$
\cite{GronauNippe,Godfrey,Gronau,Eichten,Gupta,FalkMehen,Isgur,Ebert}.  Such
corrections, which can be relatively large for $\cquark$ hadrons, are
considerably smaller for hadrons containing the more massive $\bquark$ quark.

Recent analyses at LEP, in which a charged pion produced at the primary event
vertex is combined with an inclusively reconstructed B meson \cite{BdstarLEP},
have measured an average $\BdstarUD$ mass in the range $5700-5730~\MeV$.  An
analysis combining a primary charged pion with a fully reconstructed B meson
\cite{BdstarALEPH} measures
$M_{\Btwostar}=(5739\pho^{+8}_{-11}\stat\pho^{+6}_{-4}\syst)~\MeV$ by performing
a fit to the mass spectrum which fixes the mass differences, widths and relative
rates of all spin states according to the predictions of
Reference~\citen{Eichten}.

The analysis presented here is based on combining a primary charged pion with an
inclusively reconstructed B meson.  We use new techniques both to improve the
resolution of the reconstructed $\Bmeson\pi$ mass spectrum and to unfold this
resolution from the signal components.  As a result, we are able to extract
measurements for the masses and widths of both the S-wave $\Bonestar$ decays and
the D-wave $\Btwostar$ decays.


\section{Event Selection and Reconstruction \label{sec:EventSelection}}

\subsection{Selection of \boldmath $\Zboson\rightarrow{\bquark}\bar{\bquark}$
  decays \label{sec:ZbbSelection}}

The data, collected by the L3 detector \cite{L3} in 1994 and 1995, correspond to
an integrated luminosity of $90~\pb$ with centre-of-mass energies around the Z
mass.  Hadronic Z decays are selected by making use of their characteristic
energy distributions and high multiplicity \cite{Hadrons}.  In addition, all
events are required to have an event thrust axis direction satisfying
$|\cos\theta| < 0.74$, where $\theta$ is the polar angle; to contain a primary
vertex reconstructed in three dimensions; to contain at least two calorimetric
jets, each with energy greater than $10~\GeV$; and to pass stringent detector
quality criteria for the vertexing, tracking and calorimetry.  A total of
$1\,248\,350$ events pass this selection.  Cutting on a
$\Zboson\rightarrow{\bquark}\bar{\bquark}$ event discriminant based on the
lifetime information of charged constituents \cite{bbtag} yields a
$\bquark$-enriched sample of $176\,980$ events.

A sample of 6 million simulated hadronic Z decays have been generated with
JETSET~7.4 \cite{JETSET} and passed through the L3 simulation program
\cite{Geant} to study the content of the selected data.  From this study, the
purity of the $\Zboson\rightarrow{\bquark}\bar{\bquark}$ candidates is
determined to be $83\%$ and corresponds to a selection efficiency of $65\%$.

\subsection{Selection and Reconstruction of \boldmath
            $\Bdstar\rightarrow\BorBstar\pi$ decays
            \label{sec:BpiSelection}}

The primary event vertex and secondary decay vertices are reconstructed in three
dimensions on an event-by-event basis such that each charged track can be a
constituent of no more than one vertex.  A calorimetric jet is selected for
analysis as a B candidate if it is one of the two most energetic jets in the
event, if a secondary decay vertex is reconstructed from tracks associated with
that jet, and if the distance of that vertex with respect to the primary event
vertex is greater than three times the estimated error of the measurement.

\begin{figure}[htbp]
  \setlength{\fboxrule}{2pt}
  \begin{center}
    \fbox{\epsfig{file=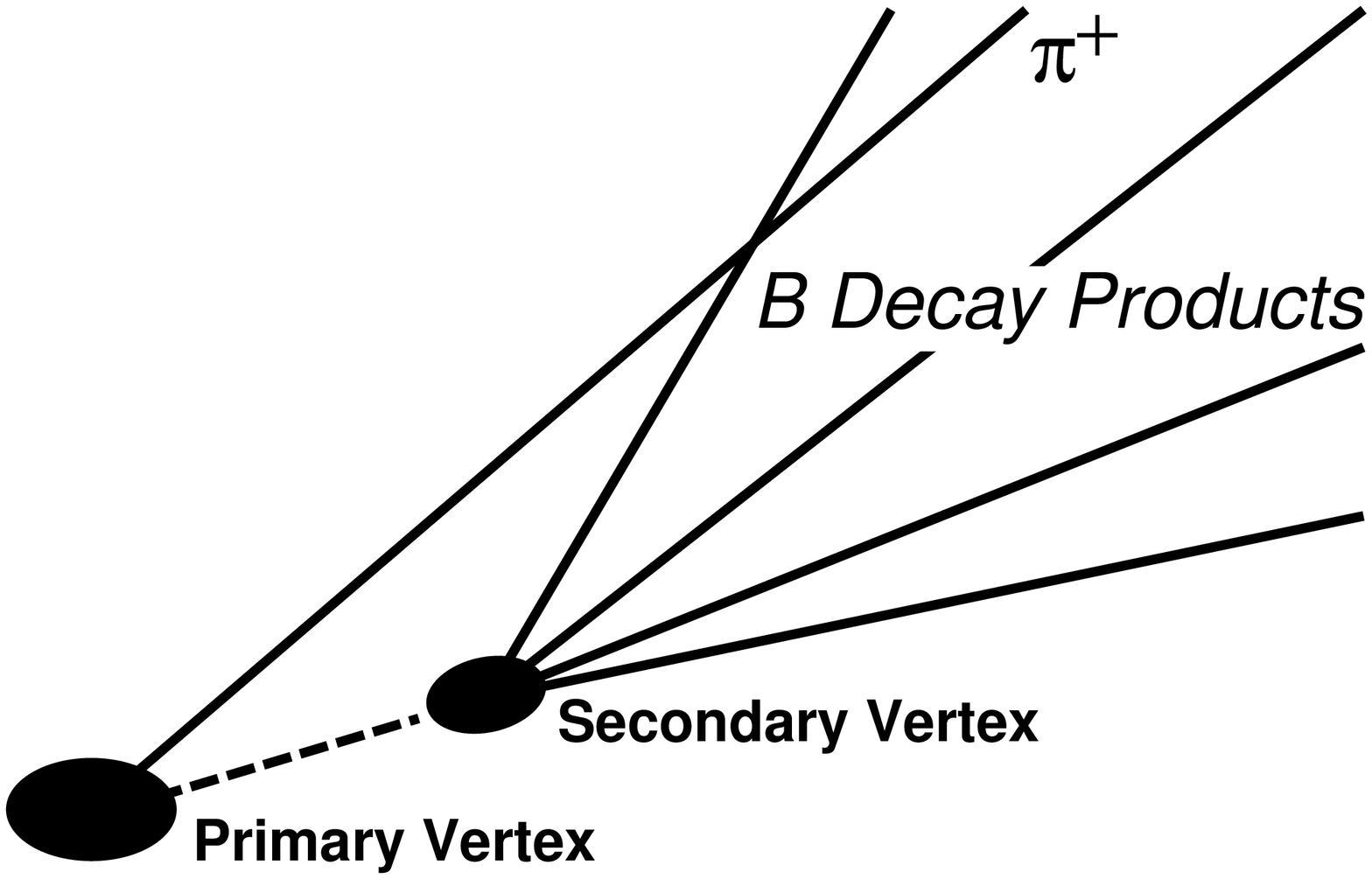,width=.8\linewidth}}
  \end{center}
  \icaption{Schematic of the decay $\Bdstar \rightarrow \BorBstar\pi$.
    Ellipses represent the reconstruction uncertainties of the vertex
    positions.  The decay pion points to the primary vertex and its
    direction is forward with respect to the direction of the B meson.
    \label{fig:BdstarDiagram}}
\end{figure}

The decay $\BdstarUD\rightarrow\BorBstar\pi$ is a strong interaction and
thus occurs at the primary event vertex.  In addition, the predicted masses for
the $l=1$ states correspond to relatively small $Q$ values, so that the decay
pion direction tends to be forward with respect to the B-meson
direction (Figure~\ref{fig:BdstarDiagram}).  We take advantage of these decay
kinematics by requiring that, for each B-meson candidate, there be at least one
track which originates from the primary event vertex and which is located in
the hemisphere defined by the jet thrust axis direction.  No attempt is made to
identify the track as a pion.  A total of $60\,205$ track-jet pairs satisfy these
criteria.

To further decrease background, which is typically due to charged particles from
fragmentation, only the track with the largest component of momentum along the
direction of the jet is selected.  This method has been found
\cite{BoscCDF,BdstarALEPH} to be an efficient means to improve the purity of the
signal.  In addition, background due to charged pions from
$\Dstar\rightarrow\Dmeson\pi$ decays is reduced by requiring the track to have a
transverse momentum with respect to the jet axis larger than $100~\MeV$.  These
selection criteria are satisfied by $48\,022$ track-jet pairs with a
$\bquark$-hadron purity of $94.2\%$.

The direction of the B candidate is estimated by taking an error-weighted
average in the $\theta$ (polar) and $\phi$ (azimuthal) coordinates
of the directions defined by the vertices and by particles with a high rapidity
relative to the jet axis.  A numerical error-propagation method
\cite{SwainTaylor} makes it possible to obtain accurate estimates for the
uncertainty of the angular coordinates measured from vertex pairs.  These errors,
as well as the error for the decay length measurement used in the secondary
vertex selection, are calculated for each pair of vertices from the associated
error matrices and determine the weight for the vertex-defined coordinate
measurements.

A second estimate for the direction of the B meson is obtained by summing the
momenta of all charged and neutral particles (excluding the decay pion
candidate) with rapidity $y>1.6$ relative to the jet axis.  Such particles have
a high probability to be decay products of the $\BorBstar$ meson.  Estimates for
the uncertainty of these coordinates are determined from simulated B-meson
decays as an average value for all events and determine the weight for the
rapidity-defined coordinate measurements.  The final B-meson direction
coordinates are taken as the error-weighted averages of the two sets of
coordinates.

The resolution for each coordinate is parametrized by a two-Gaussian
fit to the difference between the reconstructed and generated values.  For
$\theta$, the two widths are $\sigma_1=18~\mrad$ and $\sigma_2=34~\mrad$, with
$68\%$ of the B mesons in the first Gaussian.  For $\phi$, the two widths
are $\sigma_1=12~\mrad$ and $\sigma_2=34~\mrad$, with $62\%$ of the B
mesons in the first Gaussian.

The energy of the B-meson candidate is estimated by taking advantage of the
known centre-of-mass energy at LEP to constrain the measured value.  The energy
of the B meson from this method can be expressed as
\begin{equation}
  \label{eq:Benergy}
    E_{\Bmeson} = \frac{E^2_{\mathrm{cm}}-M^2_{\mathrm{recoil}}+M^2_{\Bmeson}}
                       {2E_{\mathrm{cm}}},
\end{equation}
where $E_{\mathrm{cm}}$ is the centre-of-mass energy, $M_{\mathrm{recoil}}$ is
the mass of all particles in the event with rapidity $y<1.6$, including the
decay pion candidate (regardless of its rapidity) and $M_{\Bmeson}$ is the known
B-meson mass.  The resolution is estimated by the difference between
reconstructed and generated B-meson energy values; it is best described by a
bifurcated Gaussian \cite{BifurcatedGaussian} with a lower width of $1.9~\GeV$
and an upper width of $2.8~\GeV$.


\section{Analysis of the \boldmath $\Bmeson\pi$ Mass Spectrum
         \label{sec:BpiSpectrum}}

\begin{figure}[htbp]
  \begin{center}
    \epsfig{file=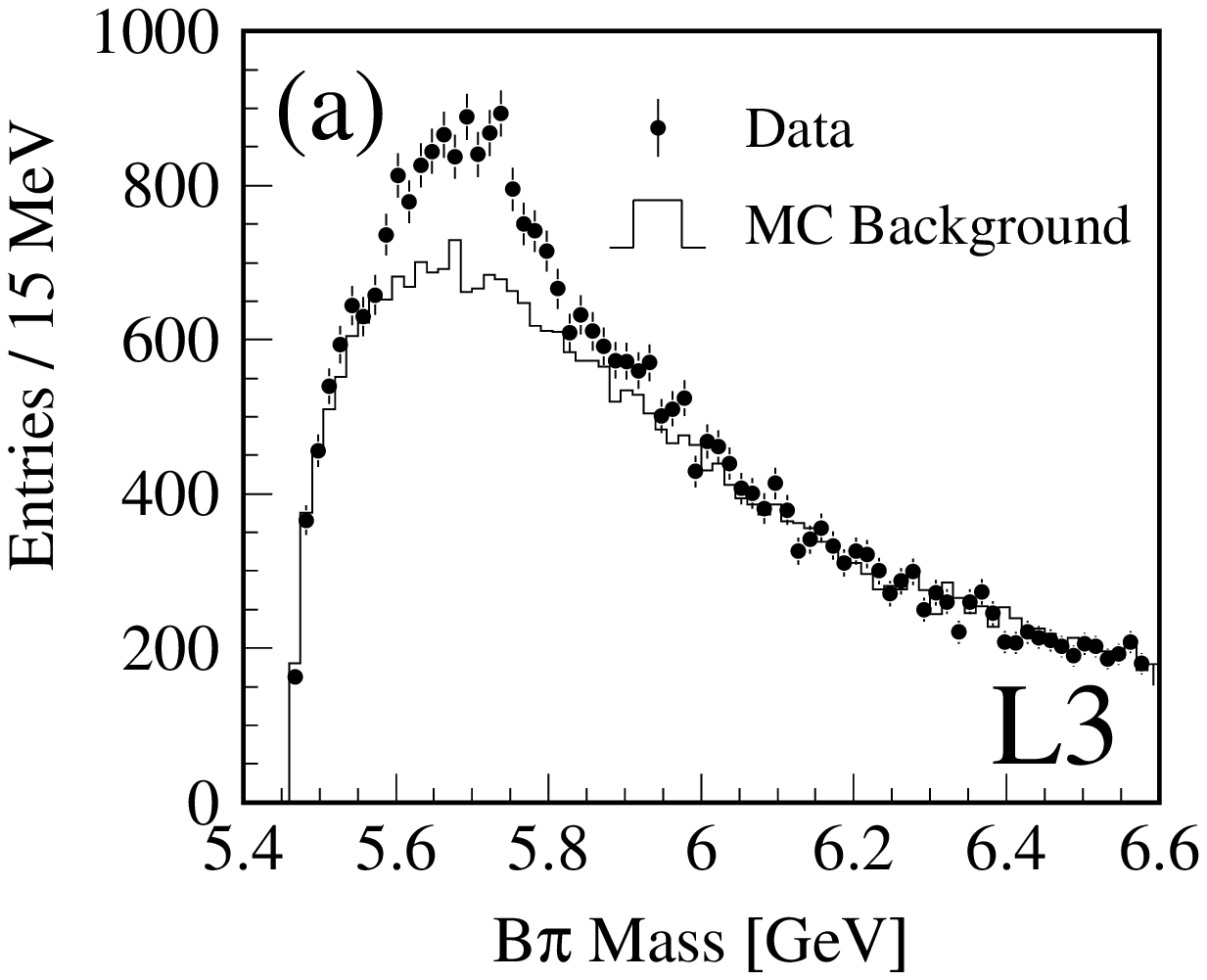,width=.8\linewidth}
  \end{center}
  \vspace*{-10mm}
  \begin{center}
    \hspace*{5mm}\epsfig{file=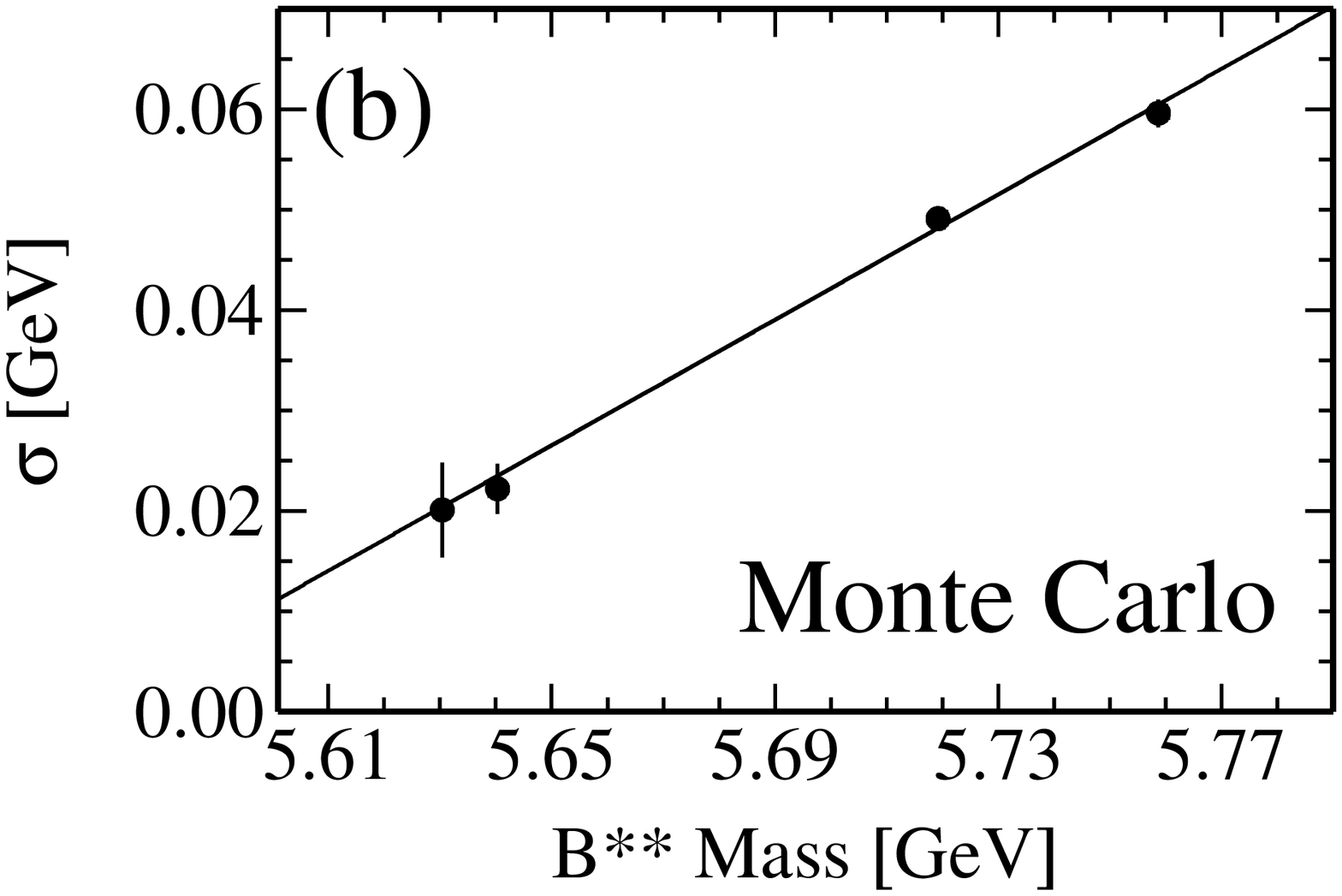,width=.8\linewidth}
  \end{center}
  \icaption{(a) Mass spectrum for selected $\Bmeson\pi$ pairs.  The points with
    error bars are data and the histogram represents the expected background
    from Monte Carlo, normalized to the sideband region $6.0-6.6~\GeV$.  (b)
    Linear fit of the extracted $\Bmeson\pi$ mass resolution for the Monte Carlo
    signal components at the generated $\Bdstar$ mass values.
    \label{fig:BpiMass}}
\end{figure}

The $\Bmeson\pi$ mass is calculated as
\begin{equation}
  \label{eq:BpiMass}
   M_{\Bmeson\pi} = \sqrt{M^2_{\Bmeson} + m^2_{\pi} + 2 E_{\Bmeson} E_{\pi} -
                          2 p_{\Bmeson} p_{\pi} \cos\alpha},
\end{equation}
where $M_{\Bmeson}$ and $m_{\pi}$ are, respectively, the known B-meson and
charged pion masses, $E_{\Bmeson}$ is the B-meson energy described in the
previous section, $p_{\pi}$ is the measured pion momentum and $\alpha$ is the
measured angle between the B-meson candidate and the decay-pion candidate.  The
resulting mass spectrum is presented in Figure~\ref{fig:BpiMass}a, along with the
Monte Carlo background, normalized to the region $6.0-6.6~\GeV$.

The background $\Bmeson\pi$ mass distribution is estimated from the Monte Carlo
sample, excluding $\Bdstar\rightarrow\BorBstar\pi$ decays, and fit with a
six-parameter threshold function given by
\begin{equation}
  \label{eq:BgdEqn}
  p_1(M_{\Bmeson\pi}-p_2)^{p_3}\:e^{\,p_4(M_{\Bmeson\pi}-p_2)   +
                                      p_5(M_{\Bmeson\pi}-p_2)^2 +
                                      p_6(M_{\Bmeson\pi}-p_2)^3}.
\end{equation}
The shape parameters, $p_2$ through $p_6$, are fixed by the fit to the simulated
background, while the overall normalization factor, $p_1$, is unconstrained in
the fit to the data spectrum.  In this manner, the normalization uncertainty is
accounted for in the statistical error.  The uncertainty due to the background
shape is accounted for in the systematic error estimate.

To examine the underlying structure, it is necessary to account for effects due
to detector resolution.  The momentum resolution of the decay-pion candidates,
which have typical momenta of $1-3~\GeV$, is a few percent, and the angular
resolution is better than $2~\mrad$.  In this case, the dominant sources of
uncertainty for the mass measurement are the B-meson angular and energy
resolutions.  Monte Carlo studies confirm that these two components of the mass
uncertainty are dominant and approximately equal in magnitude.  The following
analysis parametrizes the effects of these components on the measured
$\Bmeson\pi$ mass and then uses the parametrization to fold the resolution
effects into the fitting function.

The dependence of the $\Bmeson\pi$ mass resolution on the $\Bdstar$ mass is
studied by simulating $\Bdstar$ decays at four different values of mass and
width.  Each $\Bmeson\pi$ mass distribution is fit using a Voigt function, which
is a Breit-Wigner function convoluted with a Gaussian resolution function.  The
Breit-Wigner width is fixed to its generated value and its mass, normalization
and Gaussian resolution are extracted from the fit.

Each extracted $\Bmeson\pi$ mass value is found to agree with its generated mass
within the statistical error of the fit, and the corresponding efficiencies are
found to have no mass dependence.  The Gaussian resolution
(Figure~\ref{fig:BpiMass}b) is plotted for each generated $\Bdstar$ mass value,
and fit with a linear function.  This function is used to estimate the detector
resolution at the measured $\Bmeson\pi$ mass.

Agreement between data and Monte Carlo for the B-meson energy and angular
resolution is confirmed by analyzing $\Bstar\rightarrow\Bmeson\gamma$ decays
selected from the same sample of B mesons.  The photon selection for this test
is the same as that described in Reference~\citen{BstarL3}.  A $\Bstar$ meson
decays electromagnetically and thus has a negligible decay width compared to the
detector resolution.  As in the case of the $\Bmeson\pi$ mass resolution, the
B-meson energy and angular resolution are the dominant components of the
reconstructed $\Bmeson\gamma$ mass resolution.  Fits to the
$M_{\Bmeson\gamma}-M_{\Bmeson}$ spectra are performed with the combination of a
Gaussian signal and the background function described above.  For simulated
events, the Gaussian mean value is found to be
$M_{\Bmeson\gamma}-M_{\Bmeson}=(46.5\pm0.6)~\MeV$ with a width of
$(11.1\pm0.7)~\MeV$ for an input generator mass difference of $46.0~\MeV$.  For
data, the Gaussian mean value is found to be
$M_{\Bmeson\gamma}-M_{\Bmeson}=(45.1\pm0.6)~\MeV$ with a width of
$(10.7\pm0.6)~\MeV$.  Good agreement between the widths of the data and Monte
Carlo signals confirms that the B-meson energy and angular resolutions are well
understood and simulated.

According to spin-parity conservation, one expects mass peaks from five possible
$\BdstarUD$ decay modes:
$\Bzerostar\rightarrow\Bmeson\pi$,
$\Bonestar\rightarrow\Bstar\pi$,
$\Bone\rightarrow\Bstar\pi$,
$\Btwostar\rightarrow\Bstar\pi$ and
$\Btwostar\rightarrow\Bmeson\pi$.
No attempt is made to tag subsequent $\Bstar\rightarrow\Bmeson\gamma$ decays,
as the efficiency for selecting the soft photon is low.  As a consequence, the
effective $\Bmeson\pi$ masses for the three decays to $\Bstar$ mesons are shifted
down by the $M_{\Bstar}-M_{\Bmeson}=46~\MeV$ mass difference.  The distribution is
fit with five Voigt functions, with the relative production fractions determined
by spin counting rules.  Gaussian convolution of the decay widths is
taken from the resolution function at each measured mass value.

Additional physical constraints, presented in Table~\ref{tab:constraints}, are
based on predictions common to existing HQET models
\cite{Godfrey,Gronau,Eichten,Gupta,FalkMehen,Isgur,Ebert}.  In general, these
models predict the mass differences for the two $\jq=1/2$ states and for the two
$\jq=3/2$ states to be approximately equal and in the range $5-20~\MeV$.
Several of the models \cite{Gronau,Gupta} place the average mass of the
$\jq=3/2$ states above that of the $\jq=1/2$ states, while others
\cite{Isgur,Ebert} predict the opposite ``spin-orbit inversion.''  We
constrain $M_{\Bonestar}-M_{\Bzerostar}=12~\MeV$ and
$M_{\Btwostar}-M_{\Bone}=12~\MeV$, but allow the masses of the $\Bonestar$ and
$\Btwostar$ to be free to test the two opposing hypotheses.
\begin{table}[!htb]
  \begin{center}
    \begin{tabular}{|c|c|c|c|c|}
      \hline
      $\jq$ & Spin State & Production & Mass & Width \\
      \hline \hline
      $1/2$ & $\Bzerostar$ & $1/12$ & $M_{\Bonestar}-12~\MeV$ & $\Gamma_{\Bonestar}$ \\
      $1/2$ & $\Bonestar$  & $3/12$ & free                    & free                 \\
      $3/2$ & $\Bone$      & $3/12$ & $M_{\Btwostar}-12~\MeV$ & $\Gamma_{\Btwostar}$ \\
      $3/2$ & $\Btwostar$  & $5/12$ & free                    & free                 \\
      \hline
    \end{tabular}
    \icaption{Constraints applied to the relative production rates, masses and
      widths of the four $\BdstarUD$ spin states in the $\Bmeson\pi$ mass
      spectrum fit. \label{tab:constraints}}
  \end{center}
\end{table}

Predictions are made for the widths of the $\jq=3/2$ states by extrapolating
from measurements in the D meson system \cite{Ddstar}.  They are expected to be
approximately equal and about $20-25~\MeV$.  No precise predictions exist for the
widths of the $\jq=1/2$ states as there are no corresponding measurements in the
D system.  In general, however, they are also expected to be approximately equal,
although broader than those of the $\jq=3/2$ states.  We constrain
$\Gamma_{\Bzerostar}=\Gamma_{\Bonestar}$ and
$\Gamma_{\Bone}=\Gamma_{\Btwostar}$, but allow the widths of the $\Bonestar$ and
$\Btwostar$ to be free in the fit.

Simulated signal and background mass spectra are combined and then fit with the
functions and constraints described above.  Mass values and decay widths for the
$\Btwostar$ and $\Bonestar$ resonances and the overall normalization are
extracted from the fit and found to agree well with the generated values.  All
differences lie within the statistical errors.


\section{Fit Results \label{sec:FitResults}}

The $\Bmeson\pi$ mass spectrum for data is fit with the functions and constraints
described above.  A total of $2770 \pm 394$ events are in the signal region, and
the masses and widths of the $\Bonestar$ and $\Btwostar$ mesons are found to be
\begin{eqnarray*}
  M_{\Bonestar}      & = & (5682 \pm 23 \stat)~\MeV \\
  \Gamma_{\Bonestar} & = & (73   \pm 44 \stat)~\MeV \\
  M_{\Btwostar}      & = & (5771 \pm  7 \stat)~\MeV \\
  \Gamma_{\Btwostar} & = & (41   \pm 43 \stat)~\MeV.
\end{eqnarray*}
The fit has a $\chi^2$ of $81$ for $70$ degrees of freedom.  It does not
describe the data well in the region $5.9-6.0~\GeV$, where there appears to be
an excess of data events over the simulated background.  Exclusion of this
region from the fit yields consistent values for the signal and reduces the
$\chi^2$ to $57$ for $60$ degrees of freedom.

Several HQET models \cite{Gronau,Ebert,Gupta} predict the existence of radially
excited $(2S)$ B-meson states in the region $5.9-6.0~\GeV$.  In addition, an
inclusive measurement of $\Bmeson\pi\pi$ final states in Z decays
\cite{BpipiDELPHI} provides evidence for a resonance in the same mass region.
To account for the possible existence of these states in the mass spectrum, we
refit the data including a Gaussian function in the region of interest.

\begin{figure}[htbp]
  \begin{center}
    \epsfig{file=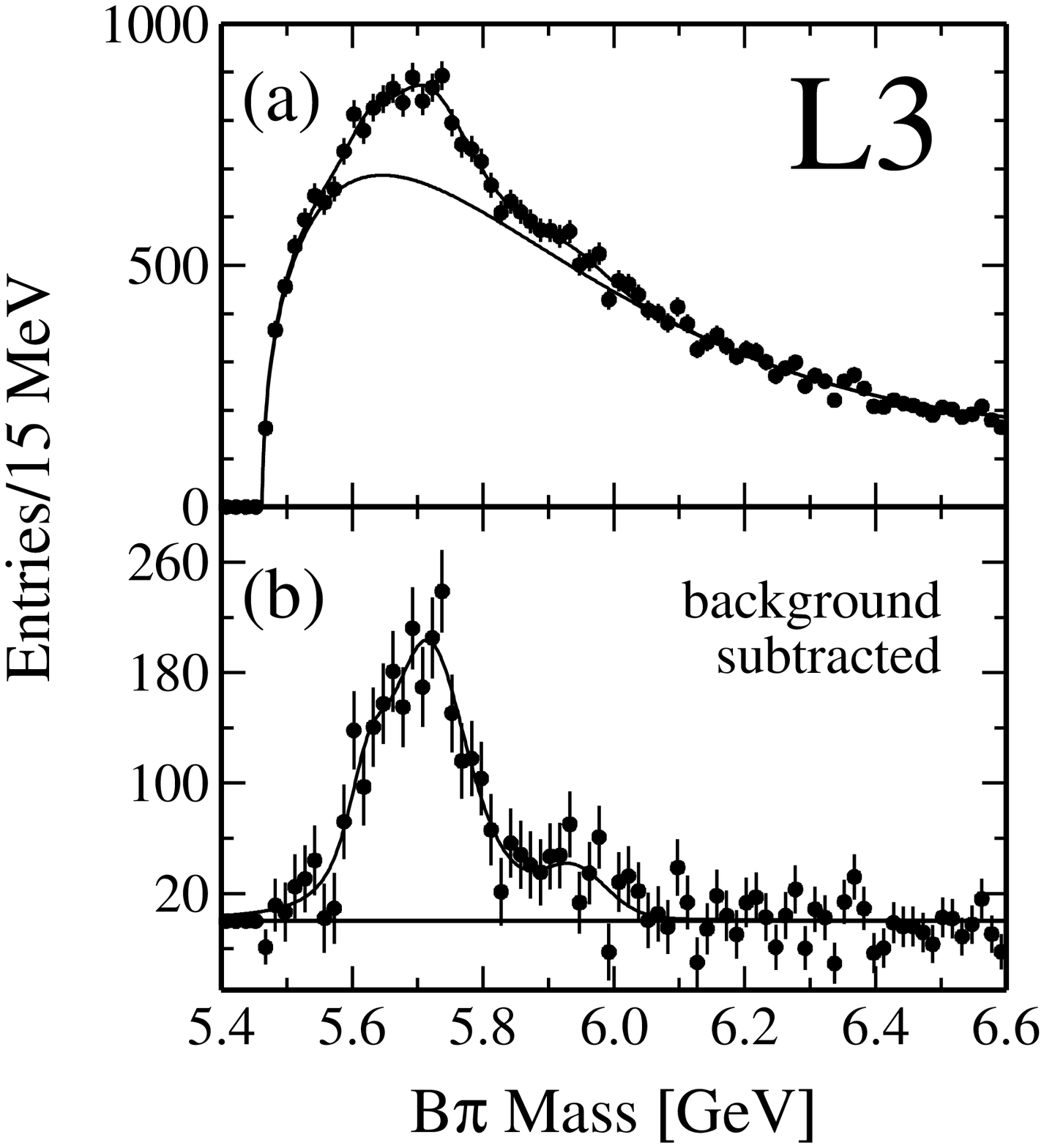,width=.9\linewidth}
  \end{center}
  \icaption{(a) Fit to the data $\Bmeson\pi$ mass distribution with the
    five-peak $\Bdstar$ signal function, the Gaussian $\Bprime$ signal function
    and the background function described in the text.  (b) The resulting
    background-subtracted distribution.
    \label{fig:Voigt}}
\end{figure}

The resulting fit, shown in Figure~\ref{fig:Voigt}, has a $\chi^2$ of $63$ for
$67$ degrees of freedom.  A total of $2784 \pm 274$ events are in the original
signal region, corresponding to the branching fraction
$\Br(\bquark\rightarrow\BdstarUD\rightarrow\BorBstar\pi)=0.32\pm0.03$.  The
masses and widths of the $\Bonestar$ and $\Btwostar$ mesons are found to be
\begin{eqnarray*}
  M_{\Bonestar}      & = & (5670 \pm 10 \stat)~\MeV \\
  \Gamma_{\Bonestar} & = & (70   \pm 21 \stat)~\MeV \\
  M_{\Btwostar}      & = & (5768 \pm  5 \stat)~\MeV \\
  \Gamma_{\Btwostar} & = & (24   \pm 19 \stat)~\MeV.
\end{eqnarray*}
In addition, a total of $297 \pm 100$ events are in the high-mass Gaussian,
denoted $\Bprime$, corresponding to the branching fraction
$\Br(\bquark\rightarrow\Bprime\rightarrow\Bmeson\pi)=0.034\pm0.011$.  The mass
and Gaussian width are found to be $M_{\Bprime}=(5936 \pm 22)~\MeV$ and
$\sigma_{\Bprime}=(50 \pm 23)~\MeV$.  Table~\ref{tab:Correlations} presents the
correlations of all free parameters in the fit.  As this fit yields the best
overall confidence level ($61.2\%$), compared to the other two fits described
above, its values are chosen as the results of the analysis.  For comparison in
the studies below, we refer to it as the ``final fit.''
\begin{table}[!htb]
  \begin{center}
    \begin{tabular}{|l|rrrrrrrrr|}
      \hline
      Param.               &
      $N_{\mathrm{bgd}}$   & $N_{\Bdstar}$   & $M_{\Bonestar}$ &
      $\Gamma_{\Bonestar}$ & $M_{\Btwostar}$ & $\Gamma_{\Btwostar}$ &
      $N_{\Bprime}$        & $M_{\Bprime}$   & $\sigma_{\Bprime}$ \\
      \hline \hline
      $N_{\mathrm{bgd}}$   & 1.000&-0.579& 0.144&-0.140& 0.100&-0.231&-0.606&-0.157&-0.432\\
      $N_{\Bdstar}$        &-0.579& 1.000& 0.004& 0.468& 0.133& 0.607& 0.050& 0.289&-0.022\\
      $M_{\Bonestar}$      & 0.144& 0.004& 1.000& 0.023& 0.276& 0.269&-0.244& 0.090&-0.190\\
      $\Gamma_{\Bonestar}$ &-0.140& 0.468& 0.023& 1.000&-0.124& 0.506&-0.115& 0.120&-0.086\\
      $M_{\Btwostar}$      & 0.100& 0.133& 0.276&-0.124& 1.000&-0.041&-0.355& 0.247&-0.362\\
      $\Gamma_{\Btwostar}$ &-0.231& 0.607& 0.269& 0.506&-0.041& 1.000&-0.127& 0.192&-0.060\\
      $N_{\Bprime}$        &-0.606& 0.050&-0.244&-0.115&-0.355&-0.127& 1.000&-0.024& 0.773\\
      $M_{\Bprime}$        &-0.157& 0.289& 0.090& 0.120& 0.247& 0.192&-0.024& 1.000& 0.058\\
      $\sigma_{\Bprime}$   &-0.432&-0.022&-0.190&-0.086&-0.362&-0.060& 0.773& 0.058& 1.000\\
      \hline
    \end{tabular}
  \end{center}
  \icaption{Table of correlations of all free parameters in the final fit.
            \label{tab:Correlations}}
\end{table}

The results of the final fit place the average mass of the $\jq=3/2$ states
$(98\pm11)~\MeV$ higher than that of the $\jq=1/2$ states.  This supports some
theoretical predictions \cite{Gronau,Gupta} but is contrary to those which
predict spin-orbit inversion \cite{Isgur,Ebert}.  To test the ability of the
procedure to discriminate between the two possible cases, additional fits are
performed.  In these fits, the widths and relative production rates of the
$\jq=1/2$ and $\jq=3/2$ states are varied, while the masses of the $\jq=1/2$
states are constrained to be either equal to those of the $\jq=3/2$ states or up
to $100~\MeV$ higher.  In all cases, the fit confidence levels decrease as the
mass of the $\jq=1/2$ states is increased relative to that of the $\jq=3/2$
states.  The highest confidence level for a fit supporting spin-orbit inversion
is found to be $8.3\%$, relative to $61.2\%$ for the final fit.

The fits performed above constrain both $M_{\Btwostar}-M_{\Bone}$ and
$M_{\Bonestar}-M_{\Bzerostar}$ to $12~\MeV$.  The results of varying this value
to test the constraint as a source of systematic uncertainty are presented
below.  It was suggested~\cite{Falk} to independently test the validity of the
$\jq=1/2$ mass-difference constraint since no corresponding measurements exist
for the D-meson system, as they do for the $\jq=3/2$ states.  This test is
performed by fitting the mass spectrum with $M_{\Bonestar}-M_{\Bzerostar}$ as an
additional free parameter.  The resulting fit, which has a $\chi^2$ of $63$ for
$66$ degrees of freedom, yields masses, widths and branching fractions for the
$\Bonestar$, $\Btwostar$ and $\Bprime$ which are consistent with the final fit.
The mass difference of the $\jq=1/2$ states is measured to be
$M_{\Bonestar}-M_{\Bzerostar}=(-39\pm71)~\MeV$.  While this value is consistent
with the original constraint, the large error indicates that the method is
insensitive to the mass difference of the $\jq=1/2$ states and, conversely, that
the results of the final fit are only weakly dependent on that constraint.

As an additional check, the mass spectrum is fit with the combination of a
single Voigt function and the background function.  The resulting fit has a
$\chi^2$ of $114$ for $76$ degrees of freedom and describes the signal region
poorly.  A total of $1854\pm153$ events are in the signal, denoted $\Bdstar$,
corresponding to the branching fraction
$\Br(\bquark\rightarrow\Bdstar\rightarrow\BorBstar\pi)=0.21\pm0.02$.  The mass
and width are found to be $M_{\Bdstar}=(5713\pm2)~\MeV$ and
$\Gamma_{\Bdstar}=(31\pm7)~\MeV$.  The hypothesis that the signal be the result
of the decay of a single resonance is highly unlikely considering the low
confidence level of the fit ($0.35\%$).

Sources of systematic uncertainty and their estimated contributions to the
errors of the measured values are summarized in Table~\ref{tab:syst}.  The
$\bquark$-hadron purity is varied from $92\%$ to $96\%$.  This variation affects
only the overall $\Bdstar$ and $\Bprime$ production branching fractions.
The contribution to these errors due to uncertainty in $\Rb$, the branching ratio
$\Br(\Zboson\rightarrow{\bquark}\bar{\bquark})
/\Br(\Zboson\rightarrow{\quark}\bar{\quark})$, is negligible.
\begin{table}[!htb]
  \begin{center}
    \begin{tabular}{|l|r|r|r|r|r|r|r|r|}
      \hline
      {Sources of} &
      {\bf $\delta M_{\Bonestar}$}    & {\bf $\delta \Gamma_{\Bonestar}$} &
      {\bf $\delta M_{\Btwostar}$}    & {\bf $\delta \Gamma_{\Btwostar}$} &
      {\bf $\delta \Br(\BdstarUD)$}   & {\bf $\delta M_{\Bprime}$}        &
      {\bf $\delta \sigma_{\Bprime}$} & {\bf $\delta \Br(\Bprime)$} \\
      {Uncertainty}
      & [$\MeV$] & [$\MeV$] & [$\MeV$] & [$\MeV$] &
      & [$\MeV$] & [$\MeV$] & \\
      \hline \hline
      $\bquark$ purity
      & ---      & ---      & ---      & ---      & $\pm 0.01$
      & ---      & ---      & $\pm 0.007$ \\
      background
      & $\pm  3$ & $\pm 19$ & $\pm  2$ & $\pm 12$ & $\pm 0.06$
      & $\pm  2$ & $\pm  2$ & $\pm 0.002$ \\
      mass constraints
      & $\pm  4$ & $\pm  7$ & $\pm  4$ & $\pm  7$ & $<0.01$
      & $\pm  1$ & $\pm  1$ & $<0.001$    \\
      width constraints
      & $<1$     & $\pm  2$ & $<1$     & $\pm  2$ & $<0.01$
      & $<1$     & $\pm  2$ & $<0.001$    \\
      fraction $\Btwostar\rightarrow\Bmeson,\Bstar$
      & $\pm  1$ & $\pm  2$ & $\pm  2$ & $\pm 4$  & $<0.01$
      & $\pm  3$ & $\pm  4$ & $\pm 0.003$ \\
      signal resolution
      & $\pm  1$ & $\pm  9$ & $\pm  2$ & $\pm  9$ & $\pm 0.01$
      & $<1$     & $<1$     & $<0.001$    \\
      signal efficiency
      & $<1$     & $\pm  1$ & $<1$     & $\pm  1$ & $\pm 0.01$
      & $<1$     & $<1$     & $\pm 0.001$ \\
      inclusion of $\Bprime$
      & $\pm 12$ & $\pm 12$ & $\pm  3$ & $\pm 17$ & $\pm 0.01$
      & ---      & ---      & ---         \\
      \hline
      {Total}
      & $\pm 13$ & $\pm 25$ & $\pm  6$ & $\pm 24$ & $\pm 0.06$
      & $\pm  4$ & $\pm  5$ & $\pm 0.008$ \\
      \hline
    \end{tabular}
    \icaption{Sources of systematic uncertainty and their estimated
              contributions to the errors of the measured values.
              Effects due to correlations between the quantities
              are taken into consideration in the total error.
              \label{tab:syst}}
  \end{center}
\end{table}

Systematic effects due to background modelling are studied by varying the shape
parameters of the background function (parameters $p_2$ through $p_6$) and by
performing the fit with other background functions.  Effects due to signal
modelling are studied by varying the constraints on the masses, widths and
relative production rates.  The mass differences $M_{\Btwostar}-M_{\Bone}$ and
$M_{\Bonestar}-M_{\Bzerostar}$ are each varied within the range $5-20~\MeV$,
covering the full range of theoretical predictions.  The ratios of the
Breit-Wigner widths $\Gamma_{\Bone}/\Gamma_{\Btwostar}$ and
$\Gamma_{\Bzerostar}/\Gamma_{\Bonestar}$ are each varied between $0.8$ and
$1.0$.  This is to account for the possibility of a slight mass dependence due
to the available phase space and the probability for
$\Bdstar\rightarrow\BorBstar\rho$ decays \cite{Quigg}.  Finally, the fraction of
$\Btwostar$ mesons decaying to $\Bstar$ mesons is varied between $1/3$ and $2/3$.

Effects due to uncertainty in the resolution function are estimated by varying
the slope and offset of the linear parametrization.  Effects due to
mass-dependent variations in the signal efficiency are estimated by performing a
linear parametrization of the Monte Carlo efficiencies, similar to that of the
resolution, and then varying the corresponding slope and offset.

$\Bmeson\pi$ pairs from the decay $\Bdstar\rightarrow\Bmeson\pi\pi$, for which
only one of the pions is tagged, are studied as a possible source of resonant
background.  The resulting reflection is found to contribute insignificantly to
the background in the low $Q$-value region.  Similarly,
$\BdstarS\rightarrow\Bmeson\Kmeson$ decays, for which the K is misidentified as
a $\pi$, are found to contribute only slightly to the low $Q$-value region, and
their effects are included in the background modelling uncertainty contribution.


\section{Conclusion \label{sec:Conclusion}}

Inclusively reconstructed B mesons in $\Zboson\rightarrow{\bquark}\bar{\bquark}$
events are combined with charged pions originating from the primary event
vertex.  A fit is performed to the $\Bmeson\pi$ mass spectrum in the framework
of Heavy Quark Symmetry \cite{HQS} and under constraints common to
several theoretical models.  The results of the fit, which provide the first
measurements of the masses and decay widths of the $\Bonestar$ ($\jq=1/2$) and
$\Btwostar$ ($\jq=3/2$) mesons, as well as the branching fraction for the
combination of $l=1$ states, are
\begin{eqnarray*}
  M_{\Bonestar}      & = & (5670 \pm   10 \stat \pm   13 \syst)~\MeV \\
  \Gamma_{\Bonestar} & = & (  70 \pm   21 \stat \pm   25 \syst)~\MeV \\
  M_{\Btwostar}      & = & (5768 \pm    5 \stat \pm    6 \syst)~\MeV \\
  \Gamma_{\Btwostar} & = & (  24 \pm   19 \stat \pm   24 \syst)~\MeV \\
  \Br(\bquark\rightarrow\BdstarUD\rightarrow\BorBstar\pi)
                     & = &  0.32 \pm 0.03 \stat \pm 0.06 \syst .
\end{eqnarray*}
In addition, an excess of events near $5.9-6.0~\GeV$ is interpreted as resulting
from the decay $\Bprime\rightarrow\BorBstar\pi$, where $\Bprime$ denotes an
excited B-meson state or mixture of states.  From the same fit, the
$\Bprime$ mass, Gaussian width and branching fraction are
\begin{eqnarray*}
  \quad M_{\Bprime}        & = & (5937 \pm     21 \stat \pm     4 \syst)~\MeV \\
  \quad \sigma_{\Bprime}   & = & (  50 \pm     22 \stat \pm     5 \syst)~\MeV \\
  \quad \Br(\bquark\rightarrow\Bprime\rightarrow\BorBstar\pi)
                           & = & 0.034 \pm 0.011 \stat \pm 0.008 \syst .
\end{eqnarray*}
For both branching fractions, isospin symmetry is employed to account for decays
to neutral pions.

These results disfavor recent theoretical models proposing spin-orbit inversion
\cite{Isgur,Ebert}, but agree well with earlier models \cite{Gronau,Gupta} and
provide support for Heavy Quark Effective Theory.


\section*{Acknowledgements}

We wish to thank Chris Quigg and Adam Falk for sharing with us their invaluable
expertise on the theoretical aspects of $\Bdstar$ production and decays.  In
addition, we wish to express our gratitude to the CERN accelerator divisions for
the excellent performance of the LEP machine.  Finally, we acknowledge the
contributions of the engineers and technicians who have participated in the
construction and maintenance of this experiment.


\newpage
\typeout{   }     
\typeout{Using author list for the B** paper ONLY!!!!!!!!!!}
\typeout{Using author list for the B** paper ONLY!!!!!!!!!!}
\typeout{Using author list for the B** paper ONLY!!!!!!!!!!}
\typeout{Using author list for the B** paper ONLY!!!!!!!!!!}
\typeout{Using author list for the B** paper ONLY!!!!!!!!!!}
\typeout{Using author list for the B** paper ONLY!!!!!!!!!!}
\typeout{$Modified: Wed Jul 28 10:29:52 1999 by clare $}
\typeout{!!!!  This should only be used with document option a4p!!!!}
\typeout{   }
%
%
%
%
%
%

\newcount\tutecount  \tutecount=0
\def\tutenum#1{\global\advance\tutecount by 1 \xdef#1{\the\tutecount}}
\def\tute#1{$^{#1}$}
\tutenum\aachen            
\tutenum\nikhef            
\tutenum\mich              
\tutenum\lapp              
\tutenum\basel             
\tutenum\lsu               
\tutenum\beijing           
\tutenum\berlin            
\tutenum\bologna           
\tutenum\tata              
\tutenum\ne                
\tutenum\bucharest         
\tutenum\budapest          
\tutenum\mit               
\tutenum\debrecen          
\tutenum\florence          
\tutenum\cern              
\tutenum\wl                
\tutenum\geneva            
\tutenum\hefei             
\tutenum\seft              
\tutenum\lausanne          
\tutenum\lecce             
\tutenum\lyon              
\tutenum\madrid            
\tutenum\milan             
\tutenum\moscow            
\tutenum\naples            
\tutenum\cyprus            
\tutenum\nymegen           
\tutenum\caltech           
\tutenum\perugia           
\tutenum\cmu               
\tutenum\prince            
\tutenum\rome              
\tutenum\peters            
\tutenum\salerno           
\tutenum\ucsd              
\tutenum\santiago          
\tutenum\sofia             
\tutenum\korea             
\tutenum\alabama           
\tutenum\utrecht           
\tutenum\purdue            
\tutenum\psinst            
\tutenum\zeuthen           
\tutenum\eth               
\tutenum\hamburg           
\tutenum\taiwan            
\tutenum\tsinghua          
{
\parskip=0pt
\noindent
{\bf The L3 Collaboration:}
\ifx\selectfont\undefined
 \baselineskip=10.8pt
 \baselineskip\baselinestretch\baselineskip
 \normalbaselineskip\baselineskip
 \ixpt
\else
 \fontsize{9}{10.8pt}\selectfont
\fi
\medskip
\tolerance=10000
\hbadness=5000
\raggedright
\hsize=162truemm\hoffset=0mm
\def\r{\rlap,}
\noindent

M.Acciarri\r\tute\milan\
P.Achard\r\tute\geneva\ 
O.Adriani\r\tute{\florence}\ 
M.Aguilar-Benitez\r\tute\madrid\ 
J.Alcaraz\r\tute\madrid\ 
G.Alemanni\r\tute\lausanne\
J.Allaby\r\tute\cern\
A.Aloisio\r\tute\naples\ 
M.G.Alviggi\r\tute\naples\
G.Ambrosi\r\tute\geneva\
% ...
\newpage
\begin{list}{A}{\itemsep=0pt plus 0pt minus 0pt\parsep=0pt plus 0pt minus 0pt
                \topsep=0pt plus 0pt minus 0pt}
\item[\aachen]
 I. Physikalisches Institut, RWTH, D-52056 Aachen, FRG$^{\S}$\\
 III. Physikalisches Institut, RWTH, D-52056 Aachen, FRG$^{\S}$
\item[\nikhef] National Institute for High Energy Physics, NIKHEF, 
     and University of Amsterdam, NL-1009 DB Amsterdam, The Netherlands
\item[\mich] University of Michigan, Ann Arbor, MI 48109, USA
\item[\lapp] Laboratoire d'Annecy-le-Vieux de Physique des Particules, 
     LAPP,IN2P3-CNRS, BP 110, F-74941 Annecy-le-Vieux CEDEX, France
\item[\basel] Institute of Physics, University of Basel, CH-4056 Basel,
     Switzerland
\item[\lsu] Louisiana State University, Baton Rouge, LA 70803, USA
\item[\beijing] Institute of High Energy Physics, IHEP, 
  100039 Beijing, China$^{\triangle}$ 
\item[\berlin] Humboldt University, D-10099 Berlin, FRG$^{\S}$
\item[\bologna] University of Bologna and INFN-Sezione di Bologna, 
     I-40126 Bologna, Italy
\item[\tata] Tata Institute of Fundamental Research, Bombay 400 005, India
\item[\ne] Northeastern University, Boston, MA 02115, USA
\item[\bucharest] Institute of Atomic Physics and University of Bucharest,
     R-76900 Bucharest, Romania
\item[\budapest] Central Research Institute for Physics of the 
     Hungarian Academy of Sciences, H-1525 Budapest 114, Hungary$^{\ddag}$
\item[\mit] Massachusetts Institute of Technology, Cambridge, MA 02139, USA
\item[\debrecen] Lajos Kossuth University-ATOMKI, H-4010 Debrecen, Hungary$^\P$
\item[\florence] INFN Sezione di Firenze and University of Florence, 
     I-50125 Florence, Italy
\item[\cern] European Laboratory for Particle Physics, CERN, 
     CH-1211 Geneva 23, Switzerland
\item[\wl] World Laboratory, FBLJA  Project, CH-1211 Geneva 23, Switzerland
\item[\geneva] University of Geneva, CH-1211 Geneva 4, Switzerland
\item[\hefei] Chinese University of Science and Technology, USTC,
      Hefei, Anhui 230 029, China$^{\triangle}$
\item[\seft] SEFT, Research Institute for High Energy Physics, P.O. Box 9,
      SF-00014 Helsinki, Finland
\item[\lausanne] University of Lausanne, CH-1015 Lausanne, Switzerland
\item[\lecce] INFN-Sezione di Lecce and Universit\'a Degli Studi di Lecce,
     I-73100 Lecce, Italy
\item[\lyon] Institut de Physique Nucl\'eaire de Lyon, 
     IN2P3-CNRS,Universit\'e Claude Bernard, 
     F-69622 Villeurbanne, France
\item[\madrid] Centro de Investigaciones Energ{\'e}ticas, 
     Medioambientales y Tecnolog{\'\i}cas, CIEMAT, E-28040 Madrid,
     Spain${\flat}$ 
\item[\milan] INFN-Sezione di Milano, I-20133 Milan, Italy
\item[\moscow] Institute of Theoretical and Experimental Physics, ITEP, 
     Moscow, Russia
\item[\naples] INFN-Sezione di Napoli and University of Naples, 
     I-80125 Naples, Italy
\item[\cyprus] Department of Natural Sciences, University of Cyprus,
     Nicosia, Cyprus
\item[\nymegen] University of Nijmegen and NIKHEF, 
     NL-6525 ED Nijmegen, The Netherlands
\item[\caltech] California Institute of Technology, Pasadena, CA 91125, USA
\item[\perugia] INFN-Sezione di Perugia and Universit\'a Degli 
     Studi di Perugia, I-06100 Perugia, Italy   
\item[\cmu] Carnegie Mellon University, Pittsburgh, PA 15213, USA
\item[\prince] Princeton University, Princeton, NJ 08544, USA
\item[\rome] INFN-Sezione di Roma and University of Rome, ``La Sapienza",
     I-00185 Rome, Italy
\item[\peters] Nuclear Physics Institute, St. Petersburg, Russia
\item[\salerno] University and INFN, Salerno, I-84100 Salerno, Italy
\item[\ucsd] University of California, San Diego, CA 92093, USA
\item[\santiago] Dept. de Fisica de Particulas Elementales, Univ. de Santiago,
     E-15706 Santiago de Compostela, Spain
\item[\sofia] Bulgarian Academy of Sciences, Central Lab.~of 
     Mechatronics and Instrumentation, BU-1113 Sofia, Bulgaria
\item[\korea] Center for High Energy Physics, Adv.~Inst.~of Sciences
     and Technology, 305-701 Taejon,~Republic~of~{Korea}
\item[\alabama] University of Alabama, Tuscaloosa, AL 35486, USA
\item[\utrecht] Utrecht University and NIKHEF, NL-3584 CB Utrecht, 
     The Netherlands
\item[\purdue] Purdue University, West Lafayette, IN 47907, USA
\item[\psinst] Paul Scherrer Institut, PSI, CH-5232 Villigen, Switzerland
\item[\zeuthen] DESY, D-15738 Zeuthen, 
     FRG
\item[\eth] Eidgen\"ossische Technische Hochschule, ETH Z\"urich,
     CH-8093 Z\"urich, Switzerland
\item[\hamburg] University of Hamburg, D-22761 Hamburg, FRG
\item[\taiwan] National Central University, Chung-Li, Taiwan, China
\item[\tsinghua] Department of Physics, National Tsing Hua University,
      Taiwan, China
\item[\S]  Supported by the German Bundesministerium 
        f\"ur Bildung, Wissenschaft, Forschung und Technologie
\item[\ddag] Supported by the Hungarian OTKA fund under contract
numbers T019181, F023259 and T024011.
\item[\P] Also supported by the Hungarian OTKA fund under contract
  numbers T22238 and T026178.
\item[$\flat$] Supported also by the Comisi\'on Interministerial de Ciencia y 
        Tecnolog{\'\i}a.
\item[$\sharp$] Also supported by CONICET and Universidad Nacional de La Plata,
        CC 67, 1900 La Plata, Argentina.
\item[$\diamondsuit$] Also supported by Panjab University, Chandigarh-160014, 
        India.
\item[$\triangle$] Supported by the National Natural Science
  Foundation of China.
\item[\dag] Deceased.
\end{list}
}
\vfill






\bibliographystyle{elsevier}


\newpage

\vspace*{50mm}

\newpage

\newpage

\end{document}